\documentclass[12pt]{iopart}
\usepackage{graphicx}
\usepackage{float}
\usepackage{bm}
\begin{document}

\bibliographystyle{apsrev}

% \preprint{Draft version, not for distribution}

%
% Title
%

\title[Fano {\it q} reversal in topological insulator Bi$_2$Se$_3$]{Fano {\it q} reversal in topological insulator Bi$_2$Se$_3$}

%
% Author list
%
%
\author{S.V. Dordevic}
\ead{dsasa@uakron.edu}%
\address{Department of Physics, The University of Akron,
Akron, Ohio 44325 USA}%
\author{G.M. Foster}
\address{Department of Physics, The University of Akron,
Akron, Ohio 44325 USA}%
\author{M.S. Wolf}
\address{Department of Physics, The University of Akron,
Akron, Ohio 44325 USA}%
\author{N. Stojilovic}
\address{Department of Physics and Astronomy, University of
Wisconsin Oshkosh, Oshkosh, Wisconsin 54901 USA}%
\author{Hechang Lei$^{\ast}$}
\address{Condensed Matter Physics and Materials Science
Department, Brookhaven National Laboratory, Upton, New York 11973 USA}%
\author{C. Petrovic}
\address{Condensed Matter Physics and Materials Science
Department, Brookhaven National Laboratory, Upton, New York 11973 USA}%
\author{Z. Chen}
\address{National High Magnetic Field Laboratory, Tallahassee,
Florida 32310 USA}%
\author{Z.Q. Li}
\address{National High Magnetic Field Laboratory, Tallahassee,
Florida 32310 USA}%
\author{L.C. Tung}
\address{Department of Physics and Astronomy,
University of North Dakota, Grand Forks, North Dakota 58202 USA}%
\address{National High Magnetic Field Laboratory, Tallahassee,
Florida 32310 USA}%

\date{\today}

%
% The abstract goes here
%
\begin{abstract}
We studied magneto-optical response of a canonical topological
insulator Bi$_2$Se$_3$ with the goal of addressing a controversial
issue of electron-phonon coupling. Magnetic-field induced
modifications of reflectance are very pronounced in the infrared
part of the spectrum, indicating strong electron-phonon coupling.
This coupling causes an asymmetric line-shape of the 60~cm$^{-1}$
phonon mode, and is analyzed within the Fano formalism. The
analysis reveals that the Fano asymmetry parameter (q) changes sign
when the cyclotron resonance is degenerate with the phonon mode. To
the best of our knowledge this is the first example of magnetic
field driven q-reversal.
\end{abstract}

%
% PACS numbers
%
%
% PACS
%
% 63.20.-e Phonons in crystal lattices
% 63.20.Dj Phonons, normal modes, phonon dispersion
% 63.20.+m Phonons in low-dimensional structures
% 71.15.Mb Density functional theory, LDA, gradient and other corrections
% 71.45.Lr CDW systems
% 74.25.-q - General properties; correlations between normal and SC states
% 74.25.Gz Optical properties
% 74.25.Kc Phonons
% 78.30.-j Infrared and Raman spectra
%

\maketitle

% \section{Introduction}

Coupling of optical phonons to charge carriers in topological
insulators is a topic of considerable current interest \cite{qi08}.
The topic is important from the fundamental point, as well as for
potential practical applications of topological insulators. The
problem has been studied with a variety of experimental techniques,
including optical spectroscopy \cite{laforge10,butch10,burch14},
Angle Resolved Photoemission Spectroscopy
\cite{hatch11,pan12,zhou13}, time-domain THz spectroscopy
\cite{armitage15} and helium scattering \cite{zhu13}. The importance
of phonons has also been studied theoretically \cite{garate}. Majority of
the measurements have been done on a canonical 3D topological
insulator Bi$_2$Se$_3$, but the results have been contradictory.
Namely, the reported values of the electron-phonon coupling
constant $\lambda$ vary in a broad range, from exceptionally weak
\cite{pan12} to moderately strong
\cite{armitage15}.

In this work we use magneto-optical spectroscopy to address this
controversial issue from another angle. Our magneto-optical results
reveal that charge carriers are indeed coupled to 60~cm$^{-1}$
optical phonon, and the coupling is causing an asymmetric line shape
of the phonon. Moreover, we find that as the magnetic field
increases, the lineshape asymmetry parameter (q) changes sign, from
negative to positive. This effect is known from other branches of
physics, and is referred to as the q-reversal. We show that the
origin of q-reversal in Bi$_2$Se$_3$ is the coupling of the phonon
to a cyclotron resonance, whose frequency increases linearly with
magnetic field.

% \section{Experimental results}

Single crystals of Bi$_2$Se$_3$ were grown at Brookhaven National
Laboratory \cite{fisk89,canfield92}.
The samples were characterized with X-ray diffraction using a
Rigaku Miniflex X-ray machine. The analysis showed that samples
were single phase, and with lattice parameters consistent with the
published values \cite{hunter98}. Samples had a typical thickness
of several millimeters and naturally flat surfaces, with surface
area of about 4~mm $\times$ 4~mm. Surfaces were cleaved before each
spectroscopic measurement.

Far-infrared magneto-reflectance ratios R($\omega$,B)/R($\omega$, 0
T) of Bi$_2$Se$_3$ were collected at the National High Magnetic
Field Laboratory in magnetic fields as high as 18 T.
Magneto-optical spectra of Bi$_2$Se$_3$ have previously been
reported only in magnetic field up to 8 T
\cite{laforge10,butch10,sushkov10,armitage11,armitage15}
and only recently in higher magnetic field \cite{orlita15}.
The reflectance ratios R($\omega$,B)/R($\omega$, 0 T) were
supplemented with the zero field data to obtain the absolute values
of reflectance in magnetic field R($\omega$, B), as the product of
the ratios with the absolute value of reflectance in zero field
\cite{dordevic05}. Zero magnetic field far-infrared measurements on
Bi$_2$Se$_3$ were performed at The University of Akron, and have
been reported previously \cite{dordevic12}. All measurements were
performed with unpolarized light, with the electric field vector
parallel to the planes.

\begin{figure}[t]%[tbp]
\vspace*{-0.5cm}%
\centerline{\includegraphics[width=9.5cm]{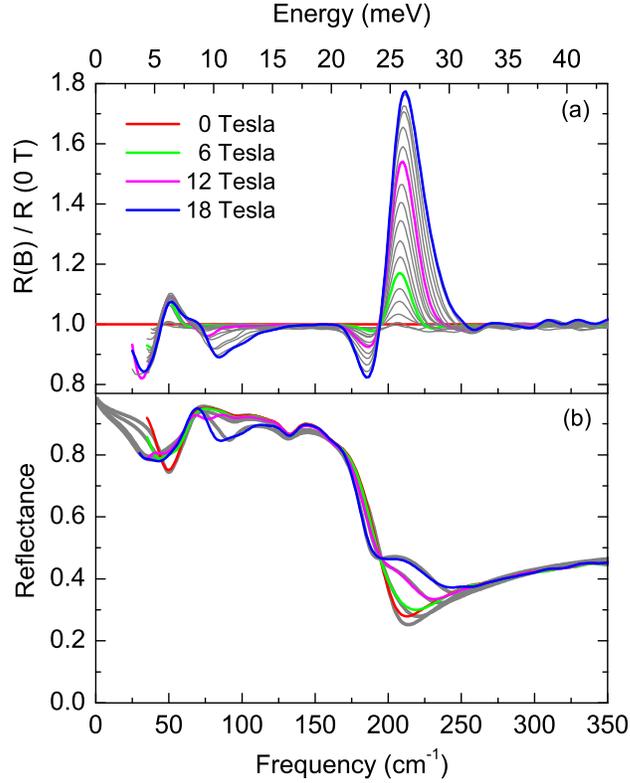}}%
\vspace*{-0.5cm}%
\caption{(Color online). (a) Magneto-reflectance ratios
R($\omega$, B)/R($\omega$, 0 T) for Bi$_2$Se$_3$ in fields up to 18
T. Very strong field induced changes are detected around the plasma
minimum, as well as around the 60~cm$^{-1}$ phonon mode. (b) The
absolute values of reflectance in magnetic field R($\omega$, B),
obtained from the ratios shown in panel (a) and the absolute values
of reflectance in zero field \cite{dordevic05,dordevic12}. Gray
lines are the best fits obtained from Eqs.~\ref{eq:dlmag} and
\ref{eq:fano}.}
\vspace*{0.0cm}%
\label{fig:ratios}
\end{figure}

Figure~\ref{fig:ratios}(a) shows the far-infrared reflectance
ratios R($\omega$, B)/R($\omega$, 0 T) in magnetic field up to 18
T, in steps of 1 T. These ratios provide the most direct evidence
for magneto-optical sensitivity of Bi$_2$Se$_3$ in the far-infrared
part of the spectrum. Large field-induced changes, as high as
75~$\%$, are detected around 220~cm$^{-1}$, much larger than
reported previously \cite{laforge10}. This is another testimony to
the excellent sample quality, which when combined with a high magnetic
field allowed for the first time \cite{laforge10,butch10,sushkov10}
the observation of a cyclotron resonance in bulk Bi$_2$Se$_3$ samples.
The cyclotron resonance in Bi$_2$Se$_3$ has so far been observed
only in thin films \cite{armitage15,orlita15}.

Figure~\ref{fig:ratios}(b) displays the absolute values of
reflectance R($\omega$,B) obtained from the ratios and the
zero-field absolute values. In zero field \cite{dordevic12}, the
most prominent feature in reflectance is the plasma edge located at
approximately 180~cm$^{-1}$. In addition, two optically active
phonon modes, observed previously
\cite{butch10,laforge10,sushkov10}, are detected in the form of
dips located at approximately 60~cm$^{-1}$ and 130~cm$^{-1}$. As
magnetic field is applied the reflectance is altered dramatically,
with the largest changes induced around the plasma minimum. In
addition, we also detect relatively large changes in the very
far-infrared part of the spectrum, around 60~cm$^{-1}$ phonon mode.

% -------------- Fig. 1 ratios and absolute R -------------
% ----------------------------------------------------------------

% ----------------------------------------------------------------
% ----------------------------------------------------------------

% \section{Discussion}

Kramers-Kronig transformation is not directly applicable to
magneto-optical spectra \cite{kuzmenko15}, so instead we apply the
fitting procedures, and from the best fits we generate the optical
functions of interest. The most frequently used model is the
Drude-Lorentz model, supplemented with the cyclotron resonance
\cite{lax67}. The dielectric function tensor
$\varepsilon_{\pm}(\omega) =
\varepsilon_{xx} \pm i \varepsilon_{xy}$ is usually represented as:

\begin{equation}
\varepsilon_{\pm}(\omega) = \varepsilon_{\infty} + \sum_{i}
\frac{\omega_{p,i}^{2}}{\omega_{0,i}^2 - \omega^2 - i \gamma_i \omega \mp
\omega_{c, i} \omega}
\label{eq:dlmag}
\end{equation}
where $\varepsilon_{\infty}$ is the high-frequency dielectric
constant, $\omega_0$ is the central frequency of the mode,
$\omega_p$ its oscillator strength and $\gamma$ its width.
$\omega_c$ is the cyclotron frequency, which is usually taken as
positive for electrons, and negative for holes. Since charge
carriers in Bi$_2$Se$_3$ have been known to be electrons
\cite{butch10}, the cyclotron frequency will be taken as positive,
and the cyclotron resonance will show up at positive frequencies in
$\varepsilon_{-}(\omega)$.

To achieve satisfactory fits and to capture the most important
features of the data a minimal model consisting of three modes is
necessary. These three modes are: 1) a Drude mode, which in
magnetic field shifts to finite frequencies and becomes a cyclotron
resonance, 2) a phonon mode at 60~cm$^{-1}$, which has been known
to be asymmetric \cite{laforge10,dordevic12}, and 3) a phonon mode
at 130~cm$^{-1}$ which does not have significant asymmetry and is
modeled as a Lorentzian \cite{dordevic12}. The cyclotron mode was
modeled with Eq.~\ref{eq:dlmag}, assuming $\omega_0$=0 for all
fields. The asymmetric 60~cm$^{-1}$ phonon was modeled as a Fano
mode \cite{fano61,kuzmenko}:

\begin{equation}
\varepsilon_{F}(\omega) =
\frac{\omega_{p,F}^{2}}{\omega_{0,F}^2 - \omega^2 - i \gamma_F \omega}
\Big( 1+i\frac{\omega_q}{\omega} \Big)^2+
\Big( \frac{\omega_{p,F} \omega_q}{\omega_{0,F} \omega}  \Big)^2
\label{eq:fano}
\end{equation}
where $\omega_{0,F}$, $\omega_{p,F}$ and $\gamma_F$ are the central
frequency, oscillator strength and width of the Fano mode,
respectively. $\omega_q$ is the Fano frequency, related to the
usual Fano asymmetry parameter q as q=$\omega_{0,F} / \omega_q$.
As discussed previously \cite{fano61}, large values of $|q|$ (i.e.
small values of $\omega_q$) indicate little or no asymmetry,
whereas small values of $|q|$ (both positive and negative) indicate
large degree of asymmetry. The form of the Fano dielectric function
given by Eq.~\ref{eq:fano} was introduced specifically for infrared
spectra \cite{kuzmenko} and has several advantages over the
conventional form \cite{fano61}. In particular, the special case of
no asymmetry (i.e. a Lorentzian lineshape) is achieved for
$\omega_q$=0, instead of $q \rightarrow \infty$.

The results of the fits at several selected fields are shown in
Fig.~\ref{fig:ratios}(b) with gray lines. As can be seen from the
plot, the model is capable of capturing all the essential features
of the data at all fields. This is significant, as the changes of
reflectance induced by magnetic field are very large and dramatic.
Namely, as the field increases, the plasma edge is gradually
altered and a characteristic "second plasma edge" develops around
220~cm$^{-1}$, similar to what was observed in Bi
\cite{laforge12} and Bi$_{1-x}$Sb$_x$ \cite{dordevic12b}. In
addition, at the lowest frequencies the reflectance is suppressed
as the cyclotron resonance enters the measured frequency window.
We note that the quality of the fits is similar at different
magnetic fields, as exemplified through $\chi^2$ measures, which
are all within 10~$\%$ \cite{comm-chi2}.

\begin{figure}[t]%[tbp]
\vspace*{-0.5cm}%
\centerline{\includegraphics[width=9.5cm]{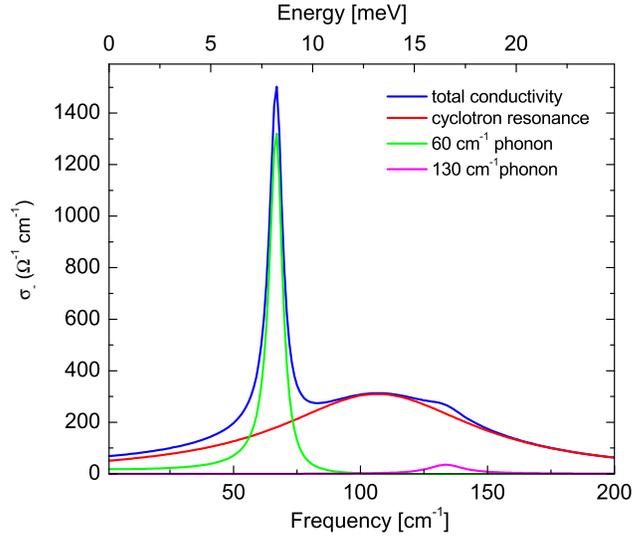}}%
\vspace*{-0.5cm}%
\caption{(Color online). Circular optical conductivity
$\sigma_-(\omega)$ generated from the best fits
(Eqs.~\ref{eq:dlmag} and \ref{eq:fano}) at 18 T is shown with blue
line. Also shown are the individual components of the fits: the
cyclotron resonance (red), the asymmetric 60~cm$^{-1}$ phonon mode
(green) and the 130~cm$^{-1}$ phonon (magenta). Note that the cyclotron
resonance is several times broader that the phonons.}
\vspace*{0.0cm}%
\label{fig:sigma}
\end{figure}

From the best fits of reflectance obtained with Eqs.~\ref{eq:dlmag}
and \ref{eq:fano} we can now generate other optical functions of
interests, in particular the circular conductivities
$\sigma_{\pm}(\omega) =  \omega (\varepsilon_{\pm}(\omega) -1)/(4 \pi i)$.
As an example in Fig.~\ref{fig:sigma} we displays the real part of
circular conductivity $\sigma_{-}(\omega)$ at 18 T. Also shown are
the individual components of the fits: the cyclotron resonance, the
asymmetric 60~cm$^{-1}$ phonon mode and the 130~cm$^{-1}$ phonon.

% -------------- Fig. 2 Sigma-  ------------------------------------
% ----------------------------------------------------------------

% ----------------------------------------------------------------
% ----------------------------------------------------------------

\begin{figure}[t]%[tbp]
\vspace*{0.0cm}%
\centerline{\includegraphics[width=9.5cm]{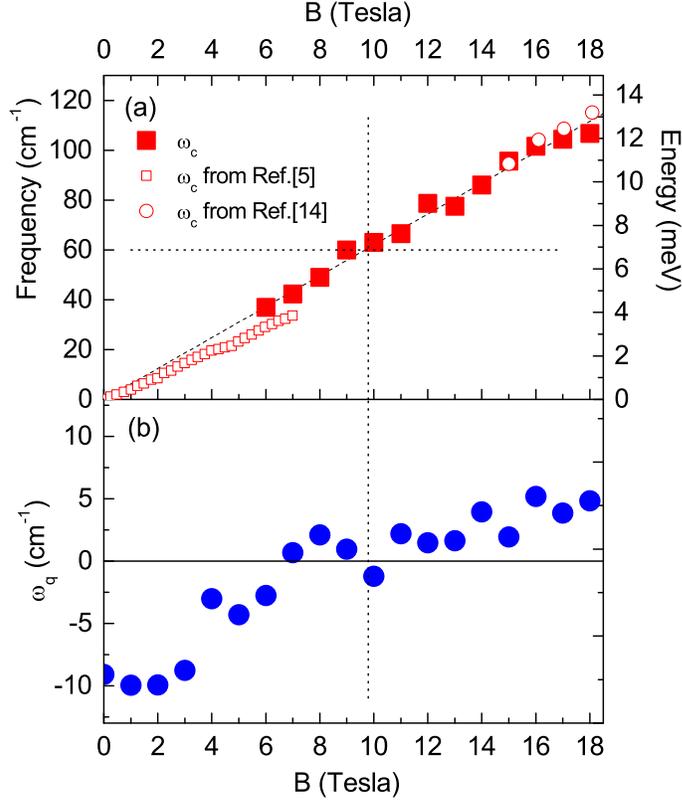}}%
\vspace*{-0.5cm}%
\caption{(Color online). Parameters of the best fits of
magneto-reflectance from Eqs.~\ref{eq:dlmag} and \ref{eq:fano}. (a)
Cyclotron resonance frequency $\omega_c$ is shown with full squares,
along with the linear fit (dashed line).
Also shown is the cyclotron frequency obtained on thin films:
open squares from Ref.~\cite{armitage15} and open circles
from Ref.~\cite{orlita15}. The horizontal dotted line represents the
frequency of the phonon. (b) Fano frequency $\omega_q$, related to
the Fano asymmetry parameter q as q=$\omega_{0,F}$/$\omega_q$. Note
that $\omega_q$ changes sign at approximately 10 Tesla (vertical
dotted line), the same field at which the cyclotron frequency is
degenerate with the phonon mode.}
\vspace*{0.0cm}%
\label{fig:parameters}
\end{figure}

Overall, the optical conductivity $\sigma_-(\omega)$ is dominated
by the phonon mode at 60~cm$^{-1}$. In zero field there is also a
Drude-like mode, but as the magnetic field increases this mode is
gradually suppressed \cite{armitage11} and shifts to finite
frequencies to become a cyclotron resonance. For fields below
approximately 5 T, the resonance is outside of our measured
frequency window ($\omega \geq $~40~cm$^{-1}$, see
Fig.~\ref{fig:ratios}), and only its tail is visible. As field
increases, the mode gradually shifts to higher frequencies, its
center frequency progressing as a linear function of the field (see
Fig.~\ref{fig:parameters}(a) below), eventually reaching
110~cm$^{-1}$ (13.6 meV) in 18 T. We point out that the
cyclotron resonance is degenerate with the 60~cm$^{-1}$ phonon mode
for magnetic field of approximately 10--11 T.

% -------------- Fig. 3 Parameters  ------------------------------------
% ----------------------------------------------------------------

% ----------------------------------------------------------------
% ----------------------------------------------------------------

The parameters of the best fits from Eqs.~\ref{eq:dlmag} and
\ref{eq:fano} are shown in Fig.~\ref{fig:parameters}.
Fig.~\ref{fig:parameters}(a) displays the cyclotron resonance
frequency $\omega_c$ (full squares). The other parameters of the
fits are not shown because they are
field independent, within the error bars of the fits.
On the other hand, the cyclotron frequency $\omega_c$ displays a
characteristic linear field dependence which extrapolates to zero
in zero field. A linear fit (shown with a dashed line) yields
$\hbar \omega_c$/B = 0.77 meV/T, from which we extract the
cyclotron effective mass m$^{*}$/m$_e$~=~0.15. This value
is in excellent agrement with the value (0.14) extracted from
a recent magnet-transmission measurement on a thick film
\cite{orlita15}. A fraction of this data are shown in
Fig.~\ref{fig:parameters}(a) with open circles. These values
of effective mass are
approximately 25~$\%$ smaller compared to the value reported
for a Bi$_2$Se$_3$ thin films \cite{armitage15}. The authors
of Ref.~\cite{armitage15} used time-domain THz spectroscopy in
magnetic fields up to 7 T and obtained the effective mass
m$^{*}$/m$_e$~=~0.20. Their data are shown in
Fig.~\ref{fig:parameters}(a) with open squares. These
recently reported values are in general agreement with
earlier reports from optical and quantum oscillations
measurements \cite{groth64,tichy79,kulbachinskii99}.

Overall the agreement between these three data sets is significant,
considering that they were obtained using different experimental
techniques. Moreover, our
measurements were done on a bulk Bi$_2$Se$_3$, whereas the other two
were performed on films (of very different thicknesses).
We also estimate carrier mobility in our sample
$\mu$=$\omega_c$/(B $\gamma$)= 1800 cm$^{-2}$/Vs, which is almost a
factor of 2 smaller compared with the value obtained on the thin film
\cite{armitage15}. However, our value is comparable with the values
obtained from Hall measurements \cite{butch10} on bulk Bi$_2$Se$_3$
samples with similar carrier density \cite{dordevic12}, and almost a
factor of 2 greater compared with the thick film \cite{orlita15}.

Fig.~\ref{fig:parameters}(b) shows the Fano frequency $\omega_q$ of
the 60~cm$^{-1}$ phonon mode and represents the most important
finding of the paper. We choose to display $\omega_q$ instead of q,
because for fields around 10 Tesla, q acquires large values
(theoretically it diverges), which renders the fitting unstable. As
the plot indicates, $\omega_q$ starts as negative
\cite{laforge10,dordevic12} for zero and small fields, but around
10 Tesla changes sign and becomes positive. This effect is referred
to as the q-reversal. As can be seen from Fig.~\ref{fig:parameters}
the asymmetry parameter changes sign when the cyclotron resonance
is degenerate with the phonon (phonon frequency is shown with a
horizontal dotted line in Fig.~\ref{fig:parameters}(a)).

Theory of the asymmetric shape of spectral lines was developed by
Fano \cite{fano61}, and subsequently used in a number of different
branches of physics. Examples of the so-called Fano resonance have
been found in studies of electron-phonon coupling in
superconductors \cite{cardona91}, scattering in photonic crystals
and plasmonic nanostructures \cite{miroshnichenko10}, atomic
photoemission spectra \cite{fano86}, and many other branches of
physics and chemistry. More recently asymmetric Fano lines have
been found in the infrared spectra of topological insulator
Bi$_2$Se$_3$ \cite{laforge10}, a few layer graphene \cite{li12} and
absorption lines of autoionizing helium \cite{ott13}, to name a
few.

The degree of lineshape asymmetry is quantified through the Fano
parameter q (or alternatively $\omega_q$). In some cases it was
observed that, as a results of some external stimulus, the Fano
parameter changes sign. This effect is called q-reversal and has
been observed for example in atomic physics \cite{connerade85},
quantum dots \cite{kobayashi02}, photonics \cite{driessen07} and
quantum waveguides \cite{klaiman07}. However, Fano q-reversals
could not be observed in the previous magneto-optical studies of
Bi$_2$Se$_3$ \cite{laforge10,armitage15} due to the limited values
of magnetic field (B~$\leq$~8~T). The observation of reversal in
this study was enabled by large magnetic fields available at the
National High Magnetic Field Laboratory. Moreover, we can trace the
origin of reversal to the field dependence of the cyclotron
resonance (Fig.~\ref{fig:parameters}).

% \section{Summary}

In summary, our magneto-optical results on Bi$_2$Se$_3$ in magnetic
fields up to 18 T reveal that electron-phonon coupling is
significant and is causing asymmetric (Fano-like) shape of the
60~cm$^{-1}$ phonon. The asymmetry parameter (q) of the phonon is a
function of magnetic field, and around 10 T changes sign, exactly
at the field at which the cyclotron resonance is degenerate with
the phonon. To the best of our knowledge, this is the first example
of Fano q-reversal induced by externally applied magnetic field.

% \section{acknowledgments}

The authors thank A.B. Kuzmenko for useful discussions. S.V.D.
acknowledges the support from The University of Akron FRG. N.S. was
supported with UW Oshkosh FDM262 grant. Work at Brookhaven is
supported by the US DOE under Contract No. DE-SC00112704 (H.L.
and C.P.). Magneto-optical measurements were carried out at the
National High Magnetic Field Laboratory, which is supported by NSF
Cooperative Agreement No. DMR-0654118, by the State of Florida, and
by the DOE.

$^{\ast}$ Present address: Department of Physics, Renmin University,
Beijing 100872, China

%%%%%%%%%%%%%%%%%%%%%%%%%%%%%%%%%%%%%%%%%%%%%%%%%%%%%%%%%%%%%%%%%
%
% The bibliography (BibTeX)
%
%\bibliography{scbo}

\begin{thebibliography}{10}
\bibitem{comm-chi2} The values of of $\chi^2$ for several characteristic
    fields are $\chi^2$(0 T) = 0.0521348, $\chi^2$(6 T) = 0.0473315,
    $\chi^2$(12 T) = 0.048939 and $\chi^2$(18 T) = 0.0504891.
\end{thebibliography}

% \end{references}

\end{document}